\documentclass[11pt]{iopart}

\usepackage{iopams}  

\usepackage{color}
\usepackage{hyperref}
\usepackage[usenames,dvipsnames,svgnames,table]{xcolor}
\usepackage{ulem}

\usepackage{mathrsfs} 
\usepackage{ulem} 

\newcommand{\f}[2]{\frac{#1}{#2}}
\newcommand{\kron}[2]{\delta^{#1}_{\phantom{#1}#2}}

\newcommand{\p}{\phantom}

\begin{document}

\paper{Kerr Black Holes and Nonlinear Radiation Memory }

\author{Thomas M\"adler$^{1}$
\footnote{Email:thomas.maedler@mail.udp.cl} and Jeffrey Winicour$^{2,3}$}

\address{
${}^1$ N\'ucleo de Astronom\'ia, Facultad de Ingenier\'ia,
Universidad Diego Portales, Av. Ej\'ercito 441, Santiago, Chile\\
${}^{2}$ Department of Physics and Astronomy \\
        University of Pittsburgh, Pittsburgh, PA 15260, USA\\
${}^{3}$ Max-Planck-Institut f\" ur
         Gravitationsphysik, Albert-Einstein-Institut, \\
	 14476 Golm, Germany \\
	}

\begin{abstract}

The Minkowski background intrinsic to the
Kerr-Schild version of the Kerr metric provides a definition
of a boosted spinning black hole. There are two
Kerr-Schild versions corresponding to ingoing or outgoing
principal null directions.
The two corresponding Minkowski backgrounds
and their associated boosts differ drastically. This
has an important implication for the gravitational
memory effect. A prior analysis
of the transition of a non-spinning Schwarzschild black
hole to a boosted state showed that the memory effect in the nonlinear regime agrees
with the linearised result based
upon the retarded Green function only if the final velocity corresponds
to a boost symmetry of the ingoing Minkowski
background. A boost with respect to the outgoing
Minkowski background is inconsistent with the absence
of ingoing radiation from past null infinity. We show that
this results extends to the transition of a Kerr black hole
to a boosted state and apply it to set upper and lower
bounds for the boost memory effect
resulting from the collision of two spinning black holes.

\end{abstract}



\section{Introduction}

The gravitational memory effect results in a net change
in the relative separation of distant particles. After a wave passes it is
determined by the difference between the initial and final
radiation strain measured by a gravitational wave detector.
The possibility of observable astrophysical consequences of the effect
was first studied in linearised gravity where the memory resulted from
a burst of massive particles which escaped to infinity,
as described by the retarded solution of the linearised
Einstein equations~\cite{zeld,brag}. 
In previous work~\cite{sky}, we showed that this
result could also be obtained in linearised theory
by considering the transition from
an initial state whose exterior was described by
a Schwarzschild metric at rest to a final state
whose exterior was a boosted exterior Schwarzschild metric.
This result was subsequently extended to the
nonlinear treatment of the
transition from a stationary to boosted Schwarzschild
exterior~\cite{boost,boost2}.
Here we further extend this treatment of the memory
effect to the boosted Kerr metric.

Our linearised treatment of the memory effect 
was based upon the stationary and boosted versions
of the ingoing Kerr-Schild version of the Schwarzschild metric
to describe the far field of the initial and final states. In order
to extend this result to the nonlinear
case three major differences from the linearised theory
had to be dealt with.

First, the linearised result was based upon the boost
symmetry of the unperturbed
Minkowski background. The Kerr-Schild metrics~\cite{ks1,ks2}
have the form
\begin{equation}
    g_{ab} = \eta_{ab} +2 H n_a n_b
    \label{eq:ksk}
\end{equation}
comprised of a Minkowski metric $\eta_{ab}$,
a principal null vector $n_a$ (with respect to
both $\eta_{ab}$ and $g_{ab}$) 
and a scalar field $H$. In the nonlinear case, the
time reflection symmetry of the Schwarzschild metric leads to
two different choices of  a ``Minkowski background''
$\eta_{ab}$ depending on whether
$n^a$ is chosen to be in the ingoing or
outgoing direction. The relation between
Boyer-Lindquist coordinates~\cite{BL} and the ingoing and
outgoing versions of Kerr-Schild coordinates
is described in Sec.~\ref{sec:ksk}. 

Second, no analogue of the Green function exists in
the nonlinear case to construct
a retarded solution. Instead, the retarded solution due
to the emission of radiation from an accelerated particle
is characterised by the absence of ingoing radiation from past
null infinity ${\mathcal I}^-$. The vanishing of radiation memory at 
${\mathcal I}^-$ is necessary for the absence of ingoing
radiation. This condition allows the
ingoing radiation strain, which forms the free characteristic initial data on
${\mathcal I}^-$, to be set to zero.
Otherwise, non-zero radiation memory at  ${\mathcal I}^-$ 
would require ingoing radiation. Since
an initial stationary Schwarzschild
metric has vanishing radiation strain 
at ${\mathcal I}^-$, the final boosted metric must also
have vanishing radiation strain at ${\mathcal I}^-$ if
there is no intervening ingoing radiation.

We found unexpected
differences in the boosts associated with the
ingoing and outgoing versions of the Kerr-Schild metric.
The memory effect due to the final velocity of
a Schwarzschild black hole is only correctly described by the
boost ${\mathcal B}$ associated with the Poincar{\'e} group of the
Minkowski background of the ingoing
Kerr-Schild metric. This is because ${\mathcal B}$
belongs to the preferred Lorentz subgroup of the
Bondi-Metzner-Sachs (BMS)~\cite{Sachs_BMS} asymptotic symmetry
group at ${\mathcal I}^-$ which is picked out by
the stationarity of the spacetime. This preferred  Lorentz
subgroup
maps asymptotically shear-free cross-sections of
${\mathcal I}^-$ into themselves. Such cross-sections
are non-distorted in the sense that their radiation strain vanishes.
Thus ${\mathcal B}$
does not introduce strain at ${\mathcal I}^-$ so
that the transition from a stationary to boosted
state is consistent with the lack of ingoing radiation 
from ${\mathcal I}^-$. But ${\mathcal B}$ is not a preferred
BMS Lorentz symmetry of future null infinity
${\mathcal I}^+$ where it induces
a supertranslation component of the BMS group.
As shown in~\cite{boost2}, the strain introduced
by this supertranslation results in non-zero radiation memory at
${\mathcal I}^+$. This memory effect
is in precise agreement with the linearised result based upon the
retarded Green function. 

Conversely, the boost symmetry
of the Minkowski metric associated with the outgoing version
of the Kerr-Schild metric is a preferred (supertranslation-free) 
BMS symmetry at ${\mathcal I}^+$. Consequently, it
introduces neither strain nor radiation
memory at ${\mathcal I}^+$.

Third, in the nonlinear regime the mass of the final
black hole depends upon the energy
loss carried off by gravitational waves. This couples the 
memory effect due to the escape of an unbound particle to the
Christodoulou memory effect~\cite{Christ_mem} due to 
energy loss from gravitational radiation.

In extending this approach from the boosted Schwarzschild to the boosted Kerr
metric there are further complications,
in addition to the algebraic complexity.
Unlike the Schwarzschild case, the principal null directions of the
Kerr metric are not hypersurface orthogonal. As a result, there is no natural
way to construct a null coordinate
system in order to study the asymptotic behavior at null infinity.
Here we show that there does exist a natural choice of
{\it hyperboloidal} coordinates which provide a
spacelike foliation extending asymptotically to null infinity.
These hyperboloidal hypersurfaces
are the null hypersurfaces of the Minkowski background for the Kerr-Schild
version of the Kerr metric, which we abbreviate by KSK metric. 
For the ingoing KSK metric (see Sec.~\ref{sec:scrim}), the hyperboloids
approach ${\mathcal I}^-$ and for the outgoing case
(see Sec.~\ref{sec:scrip}) they approach ${\mathcal I}^+$. The associated
coordinates lead to a straightforward Penrose compactification of
null infinity and allow an unambiguous treatment of the asymptotic
radiation strain and identification of the Poincar{\'e}
symmetries of the Minkowski background with BMS symmetries.
Such curved space hyperboloidal
hypersurfaces have been
utilised in formulating the Cauchy problem for Einstein's
equations in a manner suitable for radiation
studies~\cite{friedrich}. Their existence for the Kerr exterior was
investigated in terms of an asymptotic series expansion in~\cite{anil}
and in terms of an implicit 
analytic scheme for the Teukolsky equation in~\cite{anil2}.
Here we construct a simple, geometrically natural and purely analytic
hyperboloidal foliation that globally covers the entire Kerr exterior. 

An additional factor is that the Kerr metric does not have
the time reflection symmetry of the Schwarzschild
metric, but instead a $(\tau,\varphi)\rightarrow (-\tau,-\varphi)$ symmetry
in Boyer-Lindquist coordinates, as described in Sec.~\ref{sec:ksk}.
This further complicates relating asymptotic properties
at ${\mathcal I}^-$ and ${\mathcal I}^+$. 

Although these issues introduce considerable technical complication,
our main results for the Schwarzschild case extend without change to
the Kerr case. This could be expected on heuristic grounds since
the results depend only upon the asymptotic behavior of
the boosted metric for which the spin of the black
hole enters at a higher order in $1/r$ than its mass.
Nevertheless, an important aspect of the memory
effect is the supertranslation ambiguity, which enters
the bookkeeping of angular momentum. At some deeper  level,
which deserves further investigation, 
the spin of the black hole should enter.

There has been extensive elaboration and generalisation of
the memory effect since its first astrophysically relevant
discussion~\cite{zeld} in terms of the linearised description
of the burst of massive particles ejected to infinity. This 
remained the dominant mechanism until Christodoulou
demonstrated a purely nonlinear memory effect
due to mass loss from an isolated system by
gravitational waves. Extensive subsequent work
showed that the Christodoulou effect was not necessarily
a nonlinear effect but that analogous effects occurred in linear theory
due to radiative mass loss
to null infinity by Maxwell fields or other rest mass zero fields or particles.
This motivated a change in terminology of the
Christodoulou effect to a ``null'' memory effect~\cite{Bieri_cov},
as opposed to a nonlinear memory effect.
It is now understood that the memory effect results
either by the radiation to null infinity by
zero rest mass fields (or particles) or by transport to
timelike infinity by massive particles boosted with escape velocity.

Three distinct mechanisms for the memory effect
have been proposed. The first is what historically
has been called the ``linear'' memory
effect resulting from a burst of ejected particles, as described in~\cite{zeld}. For the present purpose,
we refer this as the boost mechanism. The second mechanism, i.e. null memory, generalises the Christodoulou
effect to rest mass zero fields or
particles~\cite{Bieri_em,Bieri_neutrino,tolish,Bieri_null}.
The third is a homogeneous wave mechanism~\cite{sky}
due to source free graviational waves emanating from
${\mathcal I}^-$ to ${\mathcal I}^+$.  All three
mechanisms exist in linearised theory. They can be
separated into an ``ordinary'' memory effect in which
no energy is lost to ${\mathcal I}^-$ and a null memory
effect. The treatment of null memory only depends upon
asymptotic propertiesin the neighborhood of null infinity,
which readily extends to nonlinear gravitational waves, as
demonstrated in terms of the asymptotic properties of
the vacuum Bianchi identities~\cite{frauen}, or in terms of
the asymptotic properties of Maxwell or other mass
zero radiation fields propagating in curved spacetime.
These mechanisms combine to form the net memory
effect measured by the change in radiation strain between
the infinite future and infinite past retarded times.

In this paper
we demonstrate how the memory effect due to 
a boosted Kerr black hole can be treated in
the nonlinear case. The nonlinear treatment is
purely asymptotic but, as explained above, differs considerably
from the linearised derivation based upon
the retarded flat space Green function. No linear approximations
are involved.

Supposedly the mechanism for the homogeneous
wave memory also extends to the nonlinear theory although there
is no general proof by construction of exact solutions
as in the linearised case. Other memory effects based
upon changes in angular momentum or center-of-mass
integrals have also been proposed~\cite{spin_mem,
mem_angular,nichols}. 

Although an exact Kerr metric is an
unrealistic approximation to the exterior
of a dynamical spacetime, it is a reasonable
far field approximation for the final
black hole state,
in accordance with the no hair scenario.
In Sec.~\ref{sec:mem},
we derive the nonlinear memory effect for the
transition from a stationary to boosted
Kerr black hole. In Sec.~\ref{sec:discuss},
we show how this result may be generalised
and, as an example, treat the collision of
two boosted black holes to form a final black hole. We
derive upper and lower bounds for the boost
memory in terms of the final mass of the
Kerr black hole resulting from the collision. Here
the mass of the final black hole depends upon the
energy loss in the intervening radiative period,
which leads to the null memory effect. The difference
between the final and initial radiation strains, as measured
by the net memory effect, combines both the null and
boost effects in a way which can only be determined
by knowledge of the intervening radiative period. 
This is discussed further in Sec.~\ref{sec:discuss}.

Kerr-Schild metrics have played an important
role in the construction of exact solutions~\cite{exact}.
Because their metric form (\ref{eq:ksk}) is
invariant under the Lorentz symmetry of the
Minkowski background metric $\eta_{ab}$, the boosted KSK metric
has  been important in numerical relativity in prescribing
initial data for superimposed boosted and spinning black
holes in a binary orbit~\cite{ksm1,ksm2}. The initial
data for numerical simulations are prescribed
in terms of the ingoing version of
the KSK metric, whose advanced time coordinatisation extends across
the future event horizon. The initial black hole velocities  are generated
by the boost symmetry of  the Minkowski background for the ingoing
KSK metric. This is in accord with our treatment of radiation memory.

We denote abstract spacetime indices by $a,b,...$
and coordinate indices by $\alpha,\beta,...$. In addition,
we denote 3-dimensional spatial indices for the background
inertial coordinates by $i,j,...$, and denote the associated 2-dimensional
spherical coordinates by $x^A=(\theta,\phi)$.
 We often use the standard comma notation to denote partial derivatives,
e.g.  $f_{,\alpha} =\partial f /\partial x^\alpha$.

The distinction between the ingoing and outgoing
Kerr-Schild metrics  and their associated background
Minkowski symmetries requires extra notational
care. We retain the notation in our previous papers
in which a superscript $(+)$ denotes quantities associated
with the advanced time versions of the Schwarzschild metric and a
superscript $(-)$ denotes quantities associated
with the retarded time version. Corresponding to this notation,
we use a superscript $(+)$ for quantities associated
with the ingoing version of the KSK metric and 
a superscript $(-)$ for quantities associated
with the outgoing version. As an example, the ingoing
principal null vector is denoted by $n^{(+)}_a$ and its
null rays emanate from past null infinity
${\mathcal I}^-$ and extend across the future event horizon;
and the outgoing principal
null vector is denoted by $n^{(-)}_a$, whose null
rays extend to future null infinity ${\mathcal I}^+$.
Because the memory effect is gauge invariant,
it can be computed in either in the inertial coordinates
$x^{(+)\alpha}$ or $x^{(-)\alpha}$ associated with the
ingoing or outgoing Minkowski backgrounds
of the Kerr-Schild metric,
respectively. In Sec.~\ref{sec:mem}, for technical convenience
we choose the ingoing version $x^{(+)\alpha}$.
However, the relation to the outgoing
version $x^{(-)\alpha}$ is necessary
to compute limits at ${\mathcal I}^+$.
The details of this transformation are presented in Sec.~\ref{sec:ksk}.

\section{The Kerr-Schild Kerr (KSK) metric and its associated Minkowski backgrounds}
\label{sec:ksk}

The Boyer-Lindquist coordinates~\cite{BL}, which we
denote by $ (\tau,r,\vartheta,\varphi)$, provide the
intermediate connection between the ingoing and
outgoing versions of the KSK metric.
In these coordinates, the Kerr metric is
\begin{equation}
  ds^2 = -d\tau^2 +\Sigma \Big(\frac{dr^2}{\Delta}+d\vartheta^2\Big)
    +(r^2+a^2)\sin^2\vartheta d\varphi^2 
    +\frac{2mr}{\Sigma}(a\sin^2\vartheta d\varphi -d\tau)^2  ,
    \label{eq:Schwarz_std}
\end{equation}
where $m$ is the mass, $a$ is the specific angular momentum and 
\begin{equation}
\Sigma=r^2+a^2 \cos^2\vartheta\; , \quad 
     \Delta = r^2-2mr+a^2\; .
\end{equation}
Note that this substitutes  $a\rightarrow-a$ in the 
formulae of~ \cite{ks1,ks2} to agree with the
standard convention that the sense of rotation is in
the positive $\varphi$ direction. 

The KSK metric can be expressed
in terms of either the ingoing principal null direction
$n^{(+)a}$ or the  outgoing principal null direction
$n^{(-)a}$. These two forms  
of the metric have different inertial coordinates
$x^{(\pm)\alpha} =(t^{(\pm)},x^{(\pm)},y^{(\pm)},z^{(\pm)} )$
for their corresponding Minkowski backgrounds $ \eta_{ab}^{(\pm)}$.
The main details have been worked out by considering the
$(\tau,\varphi) \rightarrow (-\tau,-\varphi)$  reflection symmetry of the
Kerr metric in Boyer-Lindquist coordinates~\cite{BL}.
The coordinate transformations leading from (\ref{eq:Schwarz_std}) to the
ingoing ($+$) or outgoing $(-)$ Kerr-Schild form involve a
generalisation of the Schwarzschild tortoise coordinate $r^*$,
\begin{eqnarray}
      r^*& =& \int \big (\frac{r^2 +a^2}{r^2-2mr+a^2}\big) dr
      \nonumber \\
      &=& r+ m \ln \big( \frac{r^2 -2mr +a^2}{4m^2} \big )
      +\frac{m^2}{\sqrt{m^2-a^2}}
      \ln\big (\frac{r-m -\sqrt{m^2-a^2}}{r-m +\sqrt{m^2-a^2}}
      \big )
      \label{eq:rstar}
\end{eqnarray}
and the intermediate angles 
\begin{equation}
\label{eq:phi_trafo}
\Phi^{\pm} = \varphi \pm a \int \f{dr}{\Delta} = 
          \varphi \pm  \frac{a}{2\sqrt{m^2-a^2}}
           \ln\Big (\frac{r-m -\sqrt{m^2-a^2}}{r-m +\sqrt{m^2-a^2}}   \Big ) \;\; .
\end{equation}
The transformation from Boyer-Lindquist to Kerr-Schild coordinates can then
be written compactly as
\begin{eqnarray}
 \quad \quad t^{(\pm)}&=& \tau \pm (r - r^*)  \label{eq:trafo_t} \\
    x^{(\pm)} +iy^{(\pm)} &=& \sqrt{r^2+a^2}\sin\vartheta
   \exp\Big\{\mathrm{i}\big[\Phi^{(\pm)}\pm \arctan(a/r)\big]\Big\}
   \label{eq:trafo_xi}\\
 \quad \quad   z^{(\pm)}&=&r\cos\vartheta .
\end{eqnarray}
Here $x^{(+)2}+y^{(+)2}=x^{(-)2}+y^{(-)2}$
so we simply denote $x^2+y^2=x^{(\pm)2}+y^{(\pm)2}$.
Similarly, we denote $z= z^{(\pm)}$.
The Boyer-Lindquist radial coordinate $r$ is 
then determined implicitly by
\begin{equation}
     \frac{x^2+y^2}{r^2+a^2} +\frac{z^2}{r^2}=1 ,
     \label{eq:r}
\end{equation}
independent of the choice of background coordinates.

The resulting Kerr Schild metrics in the
background inertial coordinates $x^{(\pm)\alpha}$ are
\begin{equation}
g_{\alpha\beta} = \eta_{\alpha\beta}^{(\pm)}
 +2Hn_\alpha^{(\pm )}n_\beta^{(\pm)}\;\;,\;\;
      H = \f{mr^3}{r^4+a^2 z^2} ,
\end{equation}
where the ingoing and outgoing versions of the principal null
vectors have components
\begin{equation}
\fl  n^{(+)}_\alpha(x^{(+)\beta}) = 
 (n^{(+)}_t,n^{(+)}_x,n^{(+)}_y,n^{(+)}_z)
  =\Big(-1, -\frac{r{x^{(+)}}+ a{y^{(+)}}}{r^2 +a^2},
     -\frac{r{y^{(+)}}- a{x^{(+)}}}{r^2 +a^2},
     - \frac{{z}}{r}\Big) ,
      \label{eq:pnd}
\end{equation}
and
\begin{equation}
\fl  n^{(-)}_\alpha(x^{(-)\beta}) = 
 (n^{(-)}_t,n^{(-)}_x,n^{(-)}_y,n^{(-)}_z)
  =\Big(-1, \frac{r{x^{(-)}}- a{y^{(-)}}}{r^2 +a^2},
     \frac{r{y^{(-)}}+ a{x^{(-)}}}{r^2 +a^2},
      \frac{{z}}{r}\Big) . 
      \label{eq:pndm}
\end{equation}
Recall that $\det(g_{\alpha\beta}) =\det(\eta^\pm_{\alpha\beta})=-1$ 
for Kerr-Schild metrics in these background coordinates.

From \eref{eq:trafo_t} and  \eref{eq:trafo_xi}, it follows that 
the  time coordinates of the two Minkowski backgrounds 
are related by
\begin{equation}
t^{(+)}=t^{(-)} +2(r-r^*)
\label{eq:ttfm}
\end{equation}
and the spatial coordinates are related by
\begin{equation}
    x^{(+)} +iy^{(+)} = \big(x^{(-)} +iy^{(-)}\big)  
     e^{i\Psi(r)} \; , \quad
  z= z^{(+)} = z^{(-)} ,
    \label{eq:stfm}
\end{equation}
where 
\begin{equation}
 \Psi(r):= \frac{a}{\sqrt{m^2-a^2}} \ln\Big (\frac{r-m -\sqrt{m^2-a^2}}{r-m 
     +\sqrt{m^2-a^2}} \Big) +2 \arctan(a/r)\;\;.
\label{eq:Psi}
\end{equation}

We set
\begin{equation}
  \rho^2 =x^2+y^2+z^2,
\end{equation}
and introduce the standard spherical coordinates
$(\rho,\theta,\phi^{(\pm)})$ for the Minkowski backgrounds,
\begin{equation}
  x^{(\pm)}+i y^{(\pm)}= \rho\sin\theta \exp[i \phi^{(\pm)}] . 
    \quad z =\rho \cos \theta \, .
    \label{eq:xypm}
\end{equation}
Here $\rho$ and $\theta$, but not $\phi^{(\pm)}$, are 
background independent.

 \section{ ${\mathcal I}^-$ and the boost symmetry}
\label{sec:scrim}

In~\cite{boost}, we showed that the linearised memory
effect arising from the retarded solution for
a freely ejected particle could be obtained from the boosted
version of the advanced time Kerr-Schild-Schwarzschild metric.
In that treatment, the boost was a Lorentz
symmetry of the linearised Minkowski background. However,
this could not  be extended unambiguously to the nonlinear
case, where there are two different choices
of boost symmetry corresponding to the Minkowski backgrounds
$ \eta^{(+)}_{\alpha\beta}$ or $ \eta^{(-)}_{\alpha\beta}$
of the ingoing  or outgoing versions of the
curved space Kerr-Schild-Schwarzschild
metric. In the curved space case, it is not the choice of ingoing or
outgoing Kerr-Schild metric (which are
algebraically equal) but the choice of boost that leads
to the essential result. In particular, because the boost symmetry
of the ingoing background $\eta^{(+)}_{\alpha\beta}$ is a preferred BMS symmetry
of ${\mathcal I}^-$ it does not produce ingoing radiation
strain at ${\mathcal I}^-$ but it does induce a supertranslation at
${\mathcal I}^+$, which
leads to outgoing radiation strain.

In the Schwarzschild case, the null hypersurfaces
determined by the principal null
directions provide a simple approach to construct null
infinity. In the Kerr case, this is more complicated because
the principal null directions are not hypersurface orthogonal.
For this reason, we describe
${\mathcal I}^-$ in the Kerr case
by considering the null spherical coordinates associated
with the Minkowski background $\eta^{(+)}_{\alpha\beta}$,
\begin{equation}
\tilde x^{(+)\alpha} = (v,\rho,\theta,\phi^{(+)}), \quad v=t^{(+)}+\rho.
\end{equation}

In these coordinates,
\begin{equation}
     \eta^{(+)}_{\alpha\beta} 
     d\tilde x^{(+)\alpha} d\tilde x^{(+)\beta} =
    - dv^2 +2dv d\rho +\rho^2 dq^{(+)2},
      \label{eq:minkplus}
\end{equation}
where
$dq^{(+)2}=q^{(+)}_{AB}dx^{(+)A}dx^{(+)B}
=d\theta^2+\sin^2 \theta d\phi^{(+)2}$ is
the unit sphere metric.
The ingoing KSK metric takes the form
\begin{equation}
     g_{\alpha\beta} 
     d \tilde x^{(+)\alpha} d \tilde x^{(+)\alpha} =
    - dv^2 +2dv d\rho+\rho^2 dq^{(+)2} +
     2H (n^{(+)}_\alpha d \tilde x^{(+)\alpha})^2,
     \label{eq:KS_hyperbolic}
\end{equation}
where, using (\ref{eq:r}) and (\ref{eq:pnd}),
\begin{equation}
   n^{(+)}_\alpha d \tilde x^{(+)\alpha} = -dv
    +(1-\frac{r}{\rho}) d\rho
   +\frac{a\rho^2\sin\theta}{r(r^2+a^2)} \big (
   a\cos\theta d\theta +r\sin\theta d\phi^{(+)}).
\end{equation}
The inverse property of Kerr-Schild metrics,
\begin{equation}
    g^{ab} = \eta^{(+)ab} -2H n^{(+)a} n^{(+)b},
\end{equation}
implies
\begin{equation}
    g^{ab}(\partial_a v)\partial_b v = 
    -2H (n^{(+)a} \partial_a v)^2.
\end{equation}
As a result, since $H\ge0$, the hypersurfaces $v=const$
are spacelike except in the limiting Schwarzschild case, where
$a=0$ and the hypersurfaces are null. 
Explicitly, (\ref{eq:pnd}) leads to
\begin{equation}
   n^{(+)\alpha} \partial_\alpha v =1-\frac{r\rho}{r^2 +a^2}
     -\frac{a^2 z^2}{r\rho(r^2 +a^2)},
\end{equation}
or, using (\ref{eq:r}),
\begin{equation}
    n^{(+)\alpha} \partial_\alpha v =1-\frac{r}{\rho}.
    \label{eq:ellv}
\end{equation}

\subsection{Compactification of past null infinity}

For $a\ne 0$, the hypersurfaces $v=const$ are
spacelike hyperboloids in the Kerr geometry
which approach ${\mathcal I}^-$ asymptotically.
In order to compactify ${\mathcal I}^-$,
we replace the hyperboloidal spherical coordinates
$\tilde x^{(+)\alpha} = (v,\rho,\theta,\phi^{(+)})$
by the compactified coordinates
$\hat x^{(+)\alpha}=(v,\ell,\theta,\phi^{(+)})$,
where $\ell=1/\rho$. In these
coordinates ${\mathcal I}^-$
is given by $\ell=0$.

Now introduce the conformally rescaled metric denoted by
$\hat g_{ab}=\ell^2 g_{ab}$. 
The  conformal metric is given in the compactified coordinates by
\begin{equation}
     \hat g_{\alpha\beta}
     d\hat x^{(+)\alpha} d \hat x^{(+)\beta} =
    - \ell^2 dv^2 -2dv d\ell + dq^{(+)2} +
     2 H(\ell n^{(+)}_\alpha d \hat x^{\alpha(+)})^2,
     \label{eq:KS_hyperbolic}
\end{equation}
with
\begin{equation}
\fl  \quad \quad  
\ell n^{(+)}_\alpha d \hat x^{(+)\alpha}
     = -\ell dv  -[\ell^{-1} - r ] d\ell
   +\frac{a \sin\theta}{r\ell(r^2+a^2)} \big (
   a\cos\theta d\theta +r\sin\theta d\phi^{(+)}).
\end{equation}

The asymptotic behaviour of
$\hat g_{\alpha\beta}$ depends upon the asymptotic expansion
of the Boyer-Lindquist coordinate $r$. From  (\ref{eq:r}),
$r$ is determined by the
quartic equation
\begin{equation}
\label{eq:r4}
\fl \quad r^4 - (x^2+y^2+z^2-a^2)r^2 -a^2z^2
=r^4 -(\rho^2 -a^2) r^2 - a^2\rho^2 \cos^2 \theta
=0. 
\end{equation}
The solution 
\begin{eqnarray}
   r &=&\f{1}{2}\sqrt{2\rho^2 -2a^2+2\sqrt{\rho^4-2\rho^2a^2
   +a^4+4a^2 z^2} }
   \label{eq:rz}\\
   &=&\f{1}{2}\sqrt{2\rho^2 -2a^2+2\sqrt{\rho^4-2\rho^2a^2
   +a^4+4a^2\rho^2 \cos^2 \theta}}
   \label{eq:rtheta} 
\end{eqnarray}
has the asymptotic $\ell$ expansion about
${\mathcal I}^-$
\begin{equation}
\label{eq:rell}
\fl \quad  r(\ell, \theta) = \ell^{-1}
 	     -\frac{a^2\sin^2\theta}{2}\;\ell
   -\frac{a^4\sin^2\theta(1-5\cos^2\theta )}{8}\; \ell^3
	     +O(\ell^{5})\; , \quad \ell =1/ \rho \, .
\end{equation}
As a result, $H$,
$\ell n^{(+)}_\alpha d \hat x^{(+)\alpha}$ and the conformal metric have the 
asymptotic $\ell$-expansions
\begin{equation}
H = m\ell
    \bigg[1 +\frac{ a^2}{2}(1-3\cos^2\theta) \ell^2 +\frac{ a^4}{8}(3-30\cos^2\theta
    +35\cos^4\theta) \ell^4+ O(\ell^6)\bigg] ,
\end{equation}
\begin{eqnarray}
  \ell n^{(+)}_\alpha d \hat x^\alpha & = &
   \ell \Big(-dv -\frac{a^2}{2}\sin^2\theta d\ell + a\sin^2\theta d\phi^{(+)}\Big)
  + \ell^2\Big(a^2\sin\theta \cos\theta d\theta \Big)\nonumber \\
 &  & 
 +\ell^3 \Big[\frac{a^4}{8}(5\cos^2\theta-1)
 \sin^2\theta d\ell 
 -a^3\sin^2\theta\cos^2\theta d\phi^{(+)}\Big]
 \nonumber\\
 &&
 +\ell^4 \Big[\frac{a^4}{2}(1-3\cos^2\theta)\sin\theta\cos\theta  d\theta\Big]
 + O(\ell^5) 
\end{eqnarray}
and
\begin{eqnarray}
\fl &&\hat g_{\alpha\beta}
d\hat x^{(+)\alpha} d \hat x^{(+)\beta}
 = -2dv d\ell + dq^{(+)2} 
    - \ell^2 dv^2   
    +2m\ell^3\Big(dv +\frac{a^2}{2}\sin^2\theta d\ell 
    - a\sin^2\theta d\phi^{(+)}\Big)^2  
\nonumber\\
\fl  &+&\ell^4 \Big[-4ma^2\Big(dv +\frac{a^2}{2}\sin^2\theta d\ell 
- a\sin^2\theta d\phi^{(+)}\Big)\sin\theta\cos\theta d\theta\Big] 
+O(\ell^5)\; ,
\label{gcrimexp}
\end{eqnarray}
with the determinant $\hat g = -\sin^2\theta$
and inverse
\begin{eqnarray}
\fl \lefteqn{ \hat g^{\alpha\beta}
\partial_{\hat x^{(+)\alpha}}
\partial_{ \hat x^{(+)\beta} }
 =-2\partial_v\partial_\ell
 + q^{(+)AB}\partial_{\hat x^{(+)A}}
    \partial_{ \hat x^{(+)B} }
    +\ell^2 \partial_\ell \partial_\ell
    -2m\ell^3\Big(
    	  \f{a^2\sin^2\theta }{2}\partial_v
	  + \partial_\ell
           + a  \partial_{\phi^{(+)}}
          \Big)^ 2
    }\nonumber\\
\fl &&   
-4a^2 m\ell^4\Big[ 
   \Big( \f{a^2\sin^2\theta }{2}\partial_v
	  + \partial_\ell
           + a  \partial_{\phi^{(+)}}\Big)\sin\theta\cos\theta\Big] \partial_\theta
+O(\ell^5) .
\end{eqnarray}

As a result, ${\mathcal I}^-$, given by $\ell=0$ with
 $\hat \nabla_a \ell |_{{\mathcal I}^-}\neq 0 $,
has Penrose compactification with metric
\begin{equation}
 \hat g_{\alpha\beta} d\hat x^{(+)\alpha} 
 d \hat x^{(+)\beta} \big |_{{\mathcal I}^-}
  = -2dv d\ell + dq^{(+)2}, 
 \end{equation}
i.e. ${\mathcal I}^-$ is a null hypersurface with standard
asymptotically Minkowskian geometry consisting of
unit sphere cross-sections.
In addition, it is straightforward to verify that
\begin{equation}
  \hat \nabla_\alpha \hat \nabla_\beta \ell |_{{\mathcal I}^-} =0,
\end{equation}
so that $\ell$ is a conformal factor in which
the shear and divergence of ${\mathcal I}^-$ vanish.
Thus $\ell$ is a preferred conformal factor
for which the compactification of ${\mathcal I}^-$
has the same asymptotic properties as described
by a conformal Bondi
frame~\cite{tam}. This allows a simple
description of the BMS asymptotic symmetries and
other physical properties of ${\mathcal I}^-$.

\subsection{Physical properties of ${\mathcal I}^-$}

The Lorentz symmetries of $\eta^{(+)}_{\alpha\beta}$
are not symmetries of the KSK metric but they
are preferred BMS symmetries of ${\mathcal I}^-$. The remaining BMS symmetries
are the supertranslations on
${\mathcal I}^-$, $v\rightarrow v + \alpha(\theta,\phi^{(+)})$.
The supertranslations with $\alpha$ composed of
$l=0$ and $l=1$ spherical harmonics correspond to
the Poincar{\'e} translations of the Minkowski background.

In the hyperboloidal coordinates $(v,\ell,x^{(+)A})$, where 
$x^{(+)A}=(\theta, \phi^{(+)})$, the
strain tensor $\sigma_{AB}(v,x^{(+)A})$ describing the 
ingoing radiation from ${\mathcal I}^-$ is determined by 
the asymptotic expansion of the metric according to
\begin{equation}
     \hat g_{AB} = q_{AB}
     +2\ell\sigma_{AB}
         +O(\ell^2) .
\end{equation}
Here $\sigma_{AB}(v,x^{(+)C})$ 
is trace-free and corresponds to the asymptotic
shear of the ingoing null hypersurfaces emanating from
the $v=const$ cross-sections of ${\mathcal I}^-$
or, equivalently, the radiation strain of the cross-sections.  It can be described
by the spin-weight-2 function 
\begin{equation}
 \sigma(v,x^{(+)A}) =  q^A q^B \sigma_{AB}(v,x^{(+)A}),
 \label{eq:sigmap}
\end{equation}
where $q^A$ is the complex polarisation dyad associated
with the unit sphere metric on ${\mathcal I}^-$, 
\begin{equation}
          q^{(+)} _{AB} = \frac{1}{2}(q_A \bar q_B +\bar q_A  q_B), \quad
           q^A \bar q_A =2, 
           \quad q^A q_A =0, \quad  q_A=q^{(+)} _{AB}q^B .
\end{equation}
For the standard choice of spherical coordinates, we set
$q^A\partial_{(+)A} =  \partial_\theta 
    +(i/\sin\theta)\partial_{\phi^{(+)} }$.
This normalisation implies
\begin{equation}
\label{ }
 \sigma=\sigma_{\theta \theta}
-\frac{\sigma_{\phi^{(+)} \phi^{(+)}} }{\sin^2\theta}
+\frac{2i\sigma_{\theta\phi^{(+)}}}{\sin\theta}\;\;,
\end{equation}
which corresponds to the standard plus/cross
decomposition, as used in~\cite{sky,boost,BSscholar}.
The
normalisation used in~\cite{boost2} inadvertently
reduced $\sigma$ by a factor of $1/2$.

In terms of the physical space description in
the associated inertial Cartesian coordinates, the
polarisation dyad
$q^A$ has components $q^i =\rho Q^i$ where
\begin{equation}
\fl \quad Q^i =\frac{1}{\rho} x^i_{,(+)A}q^A =
( \cos\theta\cos\phi^{(+)} 
-i\sin\phi^{(+)},  \cos\theta\sin\phi^{(+)}
    +i\cos\phi^{(+)},-\sin\theta)\; ,
\end{equation}    
where $Q_i \bar Q^i =2$, $Q_i  Q^i =0$ and $Q_i x^i =0$. 
Here we raise and lower the Cartesian indices $i,j,...$  for fields in
the Euclidean background according to the example $Q_i=\delta_{ij}Q^j$.  Then $\sigma$ is determined by the physical space metric
according to 
\begin{equation}
 \sigma(v,x^{(+)A}) = \lim_{\rho \rightarrow \infty}
       \frac{\rho}{2} Q^i Q^j g_{ij},
\end{equation}
where the limit at ${\mathcal I}^-$ is taken holding
$(v,x^{(+)A})$ constant.
For the unboosted KSK metric,
\begin{equation}
 \sigma(v,x^{(+)A})    = \lim_{\rho\rightarrow \infty}
 \rho  H (Q^i n^{(+)}_i )^2
    = \lim_{\rho\rightarrow \infty}
    m (Q^i n^{(+)}_i )^2.
\label{eq:sigmap}
\end{equation}
A straightforward calculation gives 
\begin{equation}\label{eq:Qiell_i}
   Q^i n^{(+)}_i = \frac {\rho a \sin\theta (a\cos\theta+ir)}{r(r^2+a^2)} .
\end{equation}

With reference to (\ref{eq:rell}), it follows that
$Q^i n^{(+)}_i=O(1/\rho)$
so that the radiation strain vanishes at the advanced
times $v=const$ picked out by the null cones of
$\eta_{\alpha\beta}^{(+)}$. However, under the 
supertranslation $v\rightarrow v+\alpha(x^{(+)A})$ the radiation 
strain has the gauge freedom (cf.~\cite{BSscholar})
\begin{equation}
  \sigma \rightarrow 
      \sigma+ q^A q^B \eth_{(+)A}\eth_{(+)B} \; 
      \alpha(x^{(+)A})
      \label{eq:super}
\end{equation}
where $\eth_{(+)A}$ is the covariant derivative with respect to
the unit sphere metric $dq^{(+)2}$. As a result,
even in a stationary
epoch where $\sigma(v,x^{(+)A})= \sigma(x^{(+)A})$,
cross-sections of ${\mathcal I}^-$  {\it distorted} by
a supertranslation have non-vanishing
strain. 

\subsection{The boost symmetry of ${\mathcal I}^-$}

The boost symmetries ${\mathcal B}$ of the Minkowski background
$\eta^{(+)}_{ab}$ are not exact symmetries of
the Kerr metric but they are asymptotic BMS symmetries of
${\mathcal I}^-$ which are preferred in the sense that
they map strain-free cross-sections into  strain-free cross-sections.
It is this asymptotic property of ${\mathcal B}$
which is essential for our treatment of the memory effect. However,
the BMS transformations are only uniquely determined to first
order in $\ell$
by the requirement that they be asymptotic symmetries.
In the compactified coordinates $\hat x^{(+)\alpha}=(v,\ell,\theta,\phi^{(+)})$, 
this leads to an equivalence class of BMS transformations in which the
identity takes the form~\cite{gw} 
\begin{equation}
\hat x^{(+)\alpha} \rightarrow  \hat x^{(+)\alpha} +O(\ell^2).
\end{equation}
Instead of the Minkowski boost
symmetries ${\mathcal B}$
we could use any equivalent BMS subgroup
in our treatment of the memory effect. The resulting
$O(\ell^2)$ terms in the boost
would not affect the radiation strain at
${\mathcal I}^-$ or ${\mathcal I}^+$.

Consider now a  Minkowski boost ${\mathcal B}$
whose 4-velocity  has components $v^\alpha = \Gamma(1, V^i)$, 
 where  $V^i =V l^i$, with direction cosines $l^i$, and
 $\Gamma =1/ \sqrt{1-V^2}$.
Under this boost,
$\eta^{(+)}_{\alpha\beta} \rightarrow \eta^{(+)}_{\alpha\beta}$.
The boosted coordinates $x^{(+)\alpha}_{\mathcal B}$ are given by
the Lorentz transformation
\begin{equation}
x^{(+)\alpha}_{\mathcal B} = \Lambda^\alpha_\beta x^{(+)\beta}\;, 
\end{equation}
where
\begin{eqnarray}
\Lambda^{t}_{\p{0}t} &=& \Gamma\;,\label{eq:LT1} \\
\Lambda^{t}_{\p{0}i} &=&\Lambda^{i}_{\p{0}t}=
 -\Gamma V_i\;,\label{eq:LT2}\\
\Lambda^{i}_{\p{0}j} &=& \kron{i}{j}+(\Gamma-1)l^i l_j\;.\label{eq:LT4}
\end{eqnarray}
The boosted coordinates are 
\begin{eqnarray}
t_{\mathcal B}^{(+)} &=& \Gamma (t^{(+)} -  V_i x^{(+)i}  ) ,\\
x^{(+)i}_{\mathcal B}&=&x^{(+)i} 
  +\Big[ -\Gamma V t^{(+)} + (\Gamma-1)l_jx^{(+)j} \Big]l^i \;,
\end{eqnarray}
e.g, for a boost in the $z$-direction, $l^x=l^y=l^z-1=0$,
$x_{\mathcal B}^{(+)}=x^{(+)}$, $y_{\mathcal B}^{(+)}=y^{(+)}$ and
\begin{eqnarray}
\label{eq:z_B}
t_B^{(+)} &=& \Gamma (t^{(+)} -  V z^{(+)}) ,\\
z^{(+)}_{\mathcal B}&=& \Gamma (z^{(+)} -  V t^{(+)}  )\; .
\end{eqnarray}
The background spherical radius $\rho$
transforms as
\begin{equation}
\label{eq:rho_B}
\fl \quad \rho^2\rightarrow \rho^2_{\mathcal B}
=x^{(+)\alpha}x^{(+)}_\alpha+(v_\alpha x^{(+)\alpha})^2 
= -[t^{(+)}]^2 + \rho^2 + \Gamma^2 ( t^{(+)} - V_i x^{(+)i})^2 \; .
\end{equation}
Setting $t^{(+)}=v -\rho$, the large $\rho$ expansion 
of $ \rho_{\mathcal B}$ about ${\mathcal I}^-$
holding $v$ constant is
\begin{equation}
\label{eq:rho_B_expa}
  \rho_{\mathcal B} =
 \rho \Gamma (1+V_i\rho^i) - \frac{v V \Gamma
      (V+l_i \rho^i)}      
       {1+V_i\rho^i}  +O(1/\rho) \;\; ,
       \label{eq:rhoexp}
\end{equation}
where
\begin{equation}
\rho^i =x^{(+)i}/\rho 
= (\sin\theta\cos\phi^{(+)},\sin\theta\sin\phi^{(+)},
\cos\theta).
\label{eq:rhoi}
\end{equation}

The boosted version of the Boyer-Lindquist radial coordinate (\ref{eq:rz}),
\begin{equation}
\label{eq:rKS_boosted}
r_{\mathcal B}=r(\rho_{\mathcal B}, z_{\mathcal B})
=\f{1}{2}\sqrt{2\rho_{\mathcal B}^2 -2a^2
+2\sqrt{\rho_{\mathcal B}^4
-2\rho_{\mathcal B}^2a^2+a^4+4a^2z^2_{\mathcal B}}}\; ,
\end{equation}
has the large $\rho$ expansion about ${\mathcal I}^-$,
holding $v=t^{(+)}+\rho$ constant,
\begin{eqnarray}
r_{\mathcal B} &=&
    (1+V_i\rho^i)\Gamma \rho
    - \frac{v V \Gamma
      (V+l_i \rho^i)}      
       {1+V_i\rho^i}  +O(1/\rho) 
       =\rho_{\mathcal B} +O(1/\rho)   .
       \label{eq:rexp}
\end{eqnarray}
This leads to the expansion of the boosted
version of the Kerr-Schild function
about ${\mathcal I}^-$,
\begin{equation}
H_{\mathcal B}= \f{mr_{\mathcal B}^3}{r_{\mathcal B}^4+a^2 z_{\mathcal B}^2}
=\frac {m} {(1+V_i\rho^i) \Gamma \rho}\bigg[ 1+
  \f{V(V+l_i\rho^i)v}
    {(1+V_i\rho^i)^2   \rho}\bigg]
    +O(1/\rho^2).
    \label{eq:Hexp}
\end{equation}
It follows from (\ref{eq:pnd}) and  (\ref{eq:rell}) that the ingoing
principle null direction has asymptotic behaviour
\begin{equation}
n^{(+)}_\alpha =-\nabla_\alpha (t^{(+)}+\rho)
+ O(1/\rho) \; .
\end{equation}
Using the covariant substitutions
$-\nabla_\alpha t\rightarrow v_\alpha$ and
$\nabla_\alpha \rho \rightarrow 
[x_\alpha+ v_\beta x^{(+)\beta}v_\alpha]/\rho_{\mathcal B} $,
its boosted version
$N_\alpha:=n_{{\mathcal B}\, \alpha}^{(+)}$
has asymptotic behaviour
\begin{equation}
N_\alpha = v_\alpha 
       - \frac {1}{\rho_{\mathcal B}}
         (x^{(+)}_\alpha+v_\beta x^{(+)\beta }v_\alpha)
       + O(1/\rho) .
       \label{eq:Lexp}
\end{equation}
Setting $t^{(+)}=v-\rho$, the expansion
of the boosted version of (\ref{eq:Qiell_i}) 
about ${\mathcal I}^-$,
holding $v$ constant, then leads to
\begin{equation}
\fl \quad Q^i N_i =(1- \frac {1}{\rho_{\mathcal B}}
      v_\beta x^{(+)\beta}) Q^i v_i   + O(1/\rho)    
      = \bigg (1-\frac{1}{\rho_{\mathcal B}}
              \rho \Gamma(1+V_i \rho^i) \bigg)Q^i v_i 
                 + O(1/\rho) \; ,
\end{equation}                 
so it follows from (\ref{eq:rhoexp}) that $Q^i N_i = O(1/\rho)$.

Thus, referring to the boosted version of
(\ref{eq:sigmap}),
\begin{equation}
 \sigma_{\mathcal B}(v,x^{(+)A})   
  = \lim_{\rho\rightarrow \infty}
 \rho  H_{\mathcal B} (Q^i N_i )^2 =0,
 \label{eq:sigmapl}
\end{equation}
i.e. the strain at ${\mathcal I}^-$ vanishes
for the boosted KSK metric, as expected
since the boost is a preferred BMS symmetry of 
${\mathcal I}^-$.

\section{Future null infinity}
\label{sec:scrip}

Following the procedure for treating ${\mathcal I}^-$,
we describe ${\mathcal I}^+$ in terms of the KSK metric
by considering retarded null spherical coordinates
associated with the Minkowski background,
$\eta^{(-)}_{ab}$,
\begin{equation}
\tilde x^{(-)\alpha} = (u,\rho,\theta,\phi^{(-)}), \quad u=t^{(-)}-\rho .
\end{equation}
In these coordinates,
\begin{equation}
     \eta^{(-)}_{\alpha \beta}
      d\tilde x^{(-)\alpha} d\tilde x^{(-)\beta} 
     =- du^2 -2du d\rho +\rho^2 dq^{(-)2} ,
      \label{eq:minkplus}
\end{equation}
where
$dq^{(-)2}=q^{(-)}_{AB}dx^{(-)A}dx^{(-)B}
=d\theta^2+\sin^2 \theta d\phi^{(-)2}$
and the outgoing version of the
KSK metric has components
\begin{equation}
     g_{\alpha \beta}
     d\tilde x^{(-)\alpha} d\tilde x^{(-)\beta} =
    - du^2 -2du d\rho +\rho^2 dq^{(-)2} +
     2H(n^{(-)}_\alpha d \tilde x^{(-)\alpha})^2.
     \label{eq:KS_hyperbolic}
\end{equation}
The inverse form of the outgoing KSK metric,
\begin{equation}
    g^{ab} = \eta^{(-)ab} -2H n^{(-)a} n^{(-)b},
\end{equation}
now implies
\begin{equation}
    g^{ab}(\partial_a u)\partial_b u = 
    -2H (n^{(-)a} \partial_a u)^2.
\end{equation}
Analogous to the ingoing case,  since $H\ge0$, the 
hypersurfaces 
$u=const$ are spacelike hyperbolae which
approach ${\mathcal I}^+$, except in the limiting 
Schwarzschild case where
they are null. 
Explicitly,  following the calculation of (\ref{eq:ellv}),
\begin{equation}
    n^{(-)\alpha} \partial_\alpha u =1-\frac{r}{\rho}.
\end{equation}

In order to compactify ${\mathcal I}^+$,
we replace the hyperboloidal spherical coordinates
$\tilde x^{(-)\alpha} = (u,\rho,\theta,\phi^{(-)})$
by the compactified
coordinates $\hat x^{(-)\alpha}=(u,\ell,\theta,\phi^{(-)})$,
where $\ell=1/\rho$, and  $\ell=0$ at ${\mathcal I}^+$.
Again we introduce the conformally rescaled
metric $\hat g_{ab}=\ell^2 g_{ab}$, 
\begin{equation}
     \hat g_{\alpha\beta}
     d \hat x^{(-)\alpha} d \hat x^{(-)\beta}=
    - \ell^2 du^2 +2du d\ell + dq^{(-)2} +
     2 H(\ell n^{(-)}_\alpha d \hat x^{\alpha(-)})^2,
     \label{eq:KS_hyperbolic}
\end{equation}
where
\begin{equation}
\fl \quad   \ell n^{(-)}_\alpha d \hat x^{(-)\alpha} 
  = -\ell du
    +[\ell^{-1} - r ] d\ell
   -\frac{a \sin\theta}{r\ell (r^2+a^2)} \big (
   a\cos\theta d\theta 
   -r\sin\theta d\phi^{(-)}). 
\end{equation}

The asymptotic behaviour of
$\hat g_{\alpha\beta}$ at ${\mathcal I}^+$ follows
from the asymptotic  $\ell$ expansion
(\ref{eq:rell}) of the
Boyer-Lindquist coordinate $r$ which leads to
\begin{eqnarray}
 \fl \quad  \ell n^{(-)}_\alpha d \hat{x}^{(-)\alpha} & = &
   \ell  \Big(-du +\frac{a^2}{2}\sin^2\theta d\ell 
   + a\sin^2\theta d\phi^{(-)}\Big)
  + \ell^2\Big(-a^2\sin\theta \cos\theta d\theta \Big)\nonumber \\
\fl &+&
 \ell^3 \Big[-\frac{a^4}{8}(5\cos^2\theta-1)
 \sin^2\theta d\ell 
     -a^3\sin^2\theta\cos^2\theta d\phi^{(-)}\Big]
 \nonumber\\
 \fl&+&
 \ell^4 \Big[-\frac{a^4}{2}(1-3\cos^2\theta)\sin\theta\cos\theta  d\theta\Big]
 + O(\ell^5) 
\end{eqnarray}
so that
\begin{eqnarray}
\fl \lefteqn{\hat g_{\alpha\beta}
    d\hat{x}^{(-)\alpha} d \hat{x}^{(-)\beta} 
 = 2du d\ell + dq^{(-)2} 
    - \ell^2 du^2  
 +2m\ell^3\Big(-du 
 +\frac{a^2}{2}\sin^2\theta d\ell 
    + a\sin^2\theta d\phi^{(-)}\Big)^2  }&&
\nonumber\\
\fl&&+\ell^4 \Big[-4ma^2\Big(-du +\frac{a^2}{2}\sin^2\theta d\ell 
+ a\sin^2\theta d\phi^{(-)}\Big)\sin\theta\cos\theta d\theta\Big] 
+O(\ell^5)\; ,
\end{eqnarray}
with determinant $\hat g = -\sin^2\theta$ and inverse 
\begin{eqnarray}
\fl \lefteqn{ \hat g^{\alpha\beta}
\partial_{\hat x^{(-)\alpha}}
\partial_{ \hat x^{(-)\beta} }
 =2\partial_u\partial_\ell
 + q^{(-)AB}\partial_{\hat x^{(-)A}}
    \partial_{ \hat x^{(-)B} }
    +\ell^2 \partial_\ell \partial_\ell
    -2m\ell^3\Big(
    	  \f{a^2\sin^2\theta }{2}\partial_u
	  - \partial_\ell
           + a  \partial_{\phi^{(-)}}
          \Big)^ 2
    }\nonumber\\
\fl &&   
+4a^2 m\ell^4\Big[ \Big( \f{a^2\sin^2\theta }{2}\partial_u
	  - \partial_\ell
           + a  \partial_{\phi^{(-)}}\Big)\sin\theta\cos\theta\Big] \partial_\theta
+O(\ell^5) .
\end{eqnarray}

We have 
\begin{equation}
  \hat g_{\alpha\beta} d\hat x^{(-)\alpha} 
  d \hat x^{(-)\beta} \big |_{{\mathcal I}^+}
  = 2du d\ell + dq^{(-)2}, 
 \end{equation}
i.e. ${\mathcal I}^+$ is a null hypersurface with standard
asymptotically Minkowskian geometry consisting of
unit sphere cross-sections.
In addition, analogous to the case for ${\mathcal I}^-$,
\begin{equation}
  \hat \nabla_\alpha \hat \nabla_\beta \ell |_{{\mathcal I}^+} =0
\end{equation}
so that $\ell$ is a preferred conformal factor in which
the shear and divergence of ${\mathcal I}^+$ vanish.

An important feature is that both ${\mathcal I}^-$
and ${\mathcal I}^+$ have universal conformal structure of
unit sphere cross-sections with the identical conformal factors
$\ell$. Moreover, (\ref{eq:Psi}) leads to the expansion\begin{equation}
   \Psi  = -2ma\ell^2-\f{8}{3}m^2 a \ell^3
   -2ma( 2m^2-a^2 \cos^2\theta) \ell^4
    + O(\ell^5) 
\end{equation}
so that $\Psi |_{\ell=0} =0$.
Consequently, (\ref{eq:stfm}) and (\ref{eq:xypm})
imply that we can set
\begin{equation}
     \phi = \phi^{(+)} |_{\ell=0}=\phi^{(-)} |_{\ell=0}
 \end{equation}
 and
\begin{equation}
dq^2=q_{AB}dx^A dx^B= d\theta^2 +\sin^2\theta d\phi^2
 = dq^{(+)2}|_{\ell=0}=dq^{(-)2} |_{\ell=0}.
\end{equation}
Thus we can use a common
unit sphere metric $q_{AB}$,  with associated
covariant derivative $\eth_A$, 
common spherical coordinates
$x^A=(\theta,\phi)$ and a common polarisation dyad
$q_{AB} =(1/2)(q_A\bar q_B+\bar q_A q_B)$  to describe
both the ingoing radiation from ${\mathcal I}^-$
and the outgoing radiation at ${\mathcal I}^+$.

\section{Boosts  and radiation memory}
\label{sec:mem}

Analogous to (\ref{eq:sigmap}), the outgoing
radiation strain at ${\mathcal I}^+$ can be described
by a spin-weight-2 function $\sigma(u,x^A)$, where
$u$ is the retarded time and $x^A=(\theta,\phi)$ are
the angular coordinates on ${\mathcal I}^+$ determined
by the Minkowski background. In these retarded
coordinates adapted to the asymptotic Minkowskian structure,
the strain is given by the
limit at ${\mathcal I}^+$, holding $u$ and $x^A$ constant,
\begin{equation}
 \sigma(u,x^A) = \lim_{\rho\rightarrow \infty}
       \frac{1}{2\rho} q^A q^B g_{AB} 
       =  q^A q^B \sigma_{AB}(u,x^A).
\end{equation}

The radiation memory $\Delta \sigma (x^C)$ at
${\mathcal I}^+$ measures
the change in the radiation strain between
infinite future and past retarded time,
\begin{equation}\label{def_memory}
       \Delta  \sigma (x^A) 
       = \sigma(u=\infty,x^A) -  \sigma(u=-\infty,x^A).
\end{equation}    
In the associated inertial Cartesian coordinates,
the dyad $q^A$ has components
$q^i = \rho Q^i$ and
\begin{equation}
 \sigma(u,x^A) = \lim_{\rho\rightarrow \infty}
       \frac{\rho}{2} Q^i Q^j g_{ij}.
       \label{eq:rstrain}
\end{equation}
For the unboosted KSK metric,
\begin{equation} \sigma(u,x^A)    = \lim_{\rho\rightarrow \infty}
    \rho H (Q^i n^{(-)}_i )^2
    = \lim_{\rho\rightarrow \infty}
    m (Q^i n^{(-)}_i )^2.
\end{equation}
A straightforward calculation gives 
\begin{equation}
   Q^i n^{(-)}_i = -\frac{\rho a^2\sin\theta\cos\theta}{r(r^2+a^2)} 
   +\frac{i\rho a\sin\theta}{r^2+a^2}. 
\end{equation}
Again using (\ref{eq:rell}), this implies $ Q^i n^{(-)}_i  =O(1/\rho)$
so that the radiation strain at ${\mathcal I}^+$ of the unboosted
KSK metric vanishes.

Consider now a system which is asymptotically
described by an unboosted Kerr metric in the retarded
past $u=-\infty$ and by a boosted KSK metric
in the future $u=\infty$, where the boost ${\mathcal B}$
is a Lorentz symmetry of $\eta^{(+)}_{\alpha\beta}$.
The radiation memory due to the boost is then given by 
\begin{equation}
     \Delta \sigma(x^A) = \sigma_{\mathcal B}(u=\infty, x^A)
     - \sigma(u=-\infty,x^A) ,
     \label{eq:memeff}
\end{equation}
where $\sigma_{\mathcal B}(u=\infty, x^A)$ is the
strain of the final boosted state and the intial strain
vanishes,  $\sigma(u=-\infty, x^A) =0$.
The strain $\sigma_{\mathcal B}(u=\infty, x^A)$ may be calculated using
either the ingoing or outgoing form of the KSK metric. 
It is
technically simpler to use the ingoing form since the boost
${\mathcal B}$ leaves $\eta^{(+)}_{\alpha\beta}$ unchanged. The final strain
for the boosted version of the KSK metric
$g_{{\mathcal B}\alpha\beta}$,  computed in the
same frame as the initial strain (\ref{eq:rstrain}), is then given by
\begin{equation}
 \sigma_{\mathcal B}(u=\infty,x^A)  
 =\lim_{u\rightarrow \infty}  \lim_{\rho\rightarrow \infty} 
       \frac{\rho}{2} Q^i Q^j g_{{\mathcal B}ij}  
       = \lim_{u\rightarrow \infty}  \lim_{\rho\rightarrow \infty}   
   \rho H_{\mathcal B}  (Q^i N_i)^2 ,
\end{equation}
where $N_i=n_{{\mathcal B}\, i}^{(+)}$. 
The leading terms in the $1/\rho$ expansion
of $\rho_{\mathcal B}$, $r_{\mathcal B}$ and
$H_{\mathcal B}$,
given in (\ref{eq:rhoexp}), (\ref{eq:rexp}) and
(\ref{eq:Hexp}), are unchanged when the limit
at ${\mathcal I}^+$ is taken holding $u$ constant.
Thus
\begin{equation}
 \sigma_{\mathcal B}(u=\infty,x^A)  
 =\lim_{u\rightarrow \infty}  \lim_{\rho\rightarrow \infty}       
  \frac{m\rho}{\rho_{\mathcal B}}(Q^i N_i)^2 .
\end{equation}
The key difference here is that the limit at ${\mathcal I}^+$
involves the boosted ingoing principal null direction
$N_\alpha$ whose asymptotic behaviour
(\ref{eq:Lexp}) leads to
\begin{equation}
 Q^iN_i = Q^iv_i \bigg ( 1+ \frac{ \Gamma (t^{(+)}
      -V_i x^{(+)i })} {\rho_{\mathcal B}} \bigg )
       + O(1/\rho).
       \label{eq;QL}
\end{equation}
Now, instead of holding the advanced time $v$ constant
to take the limit at ${\mathcal I}^-$, we hold
the retarded time $u$ constant
to take the limit at ${\mathcal I}^+$. Referring
to (\ref{eq:ttfm})
\begin{equation}
t^{(+)}=t^{(-)} +2(r-r^*) =u+\rho +2(r-r^*), 
\label{eq:tpu}
\end{equation}
where (\ref{eq:rstar}) leads to the expansion
\begin{equation}
 r^*-r
= 2m\ln\Big(\f{\rho}{2m}\Big) 
-\f{4m^2}{\rho} +O(1/\rho^2).
\end{equation}
From (\ref{eq:rho_B}), $\rho_{\mathcal B}$
has the asymptotic behaviour,
holding $u$ constant,
\begin{equation}   
   \frac{\rho_{\mathcal B}}{\rho} = \Gamma(1-\rho^i V_i) +O(1/\rho),
\end{equation}
where $\rho^i$ is defined in (\ref{eq:rhoi}).
As a result, since 
$\lim_{\rho\rightarrow \infty} \ln \rho/\rho =0$,
(\ref{eq:tpu}) leads to the limit, holding $u$ constant, 
\begin{equation}
     \lim_{\rho\rightarrow \infty}
      \frac{t^{(+)}}{\rho_{\mathcal B}}
     = \lim_{\rho\rightarrow \infty}
     \frac{\rho}{\rho_{\mathcal B}} 
     =\frac{1}{ \Gamma(1-\rho^i V_i)}
\end{equation}
and (\ref{eq;QL}) leads to
\begin{equation}
 \lim_{\rho\rightarrow \infty} Q^iN_i = 2 Q^i v_i .
\end{equation}
Thus
\begin{equation}
 \sigma_{\mathcal B}(u=\infty,x^A)  
  =\lim_{u\rightarrow \infty}  \lim_{\rho\rightarrow \infty}  
    \frac{4m\rho}{\rho_{\mathcal B}} (Q^i v_i)^2     
   =\frac{4m\Gamma}{(1-\rho^i V_i )} (Q^i V_i)^2 .
\end{equation}

The resulting boost memory due to the ejection of a
Kerr black hole of mass $m$ is
\begin{equation}
  \Delta \sigma = \frac{4m\Gamma}{(1-\rho^iV_i)} 
  (Q^i V_i)^2.
  \label{eq:mem1}
\end{equation}
This is identical to the nonlinear result~\cite{boost} 
and to the linearised result~\cite{brag} for the memory due to the relative boost between the
initial and final states of a Schwarzschild black hole.

\section{Discussion}
\label{sec:discuss}

We have shown that a boost ${\mathcal B}$ of the
Minkowski background $\eta^{(+)}_{ab} $
for the ingoing KSK metric
leads to a model for the memory effect
for an initially stationary Kerr black hole which, after
some accelerating and  radiating stage,
results in a final boosted state. The result is consistent with the absence
of ingoing radiation and does not depend upon any linearised approximation. The full memory effect combines the boost memory, which we now denote
by $\Delta \sigma_{\mathcal B}$, and
the null memory $\Delta \sigma_{\mathcal N}$ resulting
from radiative energy loss to ${\mathcal I}^+$.
The net memory effect is then
\begin{equation}
   \Delta \sigma= \Delta \sigma_{\mathcal B}
  + \Delta \sigma_{\mathcal N}.
\end{equation}

Although we have concentrated on the boost memory for the simple process consisting of the ejection
of a boosted Kerr black hole, the result
can be generalised. First, the asymptotic
Lorentz symmetry at null infinity implies
that the radiation memory for the transition of a black hole
from a rest
state to a boosted state with mass $m$ and velocity $V^i$
is the same as the memory for a black hole
of mass $m$ with initial
velocity $-V^i$ and zero final velocity. In addition,
even in the nonlinear theory it is expected
that the superposition principle
holds for particles at infinite separation
since the constraints vanish in that limit.
This allows the memory
effect to be generalised to a system of particles.

As a simple example, we consider the collision
of two distant Kerr black
holes of mass $m$ with
initial velocities $V^i$ and $-V^i$ in the $z$-direction
which come to rest in a final state with mass $M$.
The collision is constrained by the
Bondi mass loss fomula, which requires
\begin{equation}
   2m\Gamma -M =\int_{-\infty}^{\infty} du
     \oint  \|\partial_u \sigma\|^2 \sin\theta d\theta d\phi.
   \label{eq:eneq}
\end{equation} 
The null memory is determined nonlocally from the 
integrated radiation flux. In the case of gravitational radiation,
Christodoulou's result~\cite{Christ_mem} can be expressed
in terms of the radiation strain
according to~\cite{frauen,boost}
\begin{equation}
   \bar q^A \bar q^B \eth_A \eth_B
   \Delta \sigma_{\mathcal N} =
     \int_{-\infty}^{\infty} \|\partial_u \sigma \|^2 du
     -\Delta {\mathcal P} ,
\end{equation}
where $\Delta {\mathcal P}$ cancels the $l=0$ and $l=1$
spherical harmonics in the flux integral. The 
$l=0$ harmonic of the integrand determines the mass loss
via (\ref{eq:eneq}) and the $l=1$ harmonics determine
the momentum loss. These harmonics do not enter
the memory effect which is a spin-weight-2
quantity with $l\ge 2$.

According to (\ref{eq:mem1}), the boost memory
for this process is
\begin{equation}
  \Delta \sigma_{\mathcal B} = 4m\Gamma(Q^i V_i )^2
  \big (\frac{1}{1+\rho^i V_i }+\frac{1}{1- \rho^i V_i } \big )
  =\frac{8m \Gamma V^2\sin^2\theta}{1-V^2\cos^2 \theta}.
  \label{eq:mem}
\end{equation}
As a consequence of (\ref{eq:eneq}), $2m\Gamma -M >0$
so that the boost memory (\ref{eq:mem}) for the collision has a
lower bound determined by the mass of the final black hole,
\begin{equation}
  \Delta \sigma_{\mathcal B} > 
  \frac{ 4M V^2\sin^2\theta}{1-V^2\cos^2 \theta}.
\end{equation}
This lower bound is largest when the merger of the two black holes takes place slowly so that $M\approx 2m \Gamma$
and there is negligible radiative energy loss and
negligible null memory.

The memory effect is also constrained by Hawking's
area increase law for the event horizon in the
merger of two black holes~\cite{hawk2}.
For the above collision of initially distant Schwarzschild black holes to form a Kerr black hole, this leads to
\begin{equation}
   4m^2 < M(M+\sqrt{M^2-A^2}) =M^2(1+\cos\chi),
   \quad 0\le \chi \le \pi/2 ,
   \label{eq:areaineq}
\end{equation}
where $A=M\sin\chi$ is the specific angular momentum of
the final Kerr black hole as determined by the initial
impact parameter. Thus  (\ref{eq:areaineq}) implies
\begin{equation}
  2m <M\sqrt{1+\cos\chi}  .
  \label{eq:mineq}
  \end{equation} 
As a result, the boost memory (\ref{eq:mem}) has both
upper and lower bounds in terms of the final black hole
mass,
\begin{equation}
  \frac{4M\Gamma V^2\sin^2\theta\sqrt{1+\cos\chi}}{1-V^2\cos^2 \theta} >
  \Delta \sigma_{\mathcal B} >
   \frac{ 4M V^2\sin^2\theta}{1-V^2\cos^2 \theta}.
\end{equation}

There is no universal way to separate the net memory
effect into boost and null mechanisms without knowledge
of the strain in the intermediate radiative period. However,
there are two extremes in which the net memory takes
either the pure boost or pure null form.
The first is the adiabatic limit,
discussed in~\cite{boost}, in which the intermediate processes take place
slowly and $\Delta \sigma_{\mathcal N}$ is negligible.
This limit was first considered in~\cite{Smarr}
and is now referred to in the literature as
the infrared or ``soft graviton''
limit~\cite{pasterski,mem_soft_theorem}.
In the other extreme, no boosted
massive particles are captured or ejected from the
intrinsic rest frame of the system, so that
$ \Delta \sigma= \Delta \sigma_{\mathcal N}$.
An example would be
a binary in an initial Newtonian orbit which subsequently
inspirals to form a stationary black hole with respect to
the initial rest frame.

In~\cite{boost}, we analysed how radiation memory
affects angular momentum conservation. In a non-radiative regime where
$\partial_u \sigma=0$,
the supertranslation freedom (\ref{eq:super})
can be used to pick out preferred cross-sections
of ${\mathcal I}^+$ by setting the electric component
of $\sigma$ to 0. These
preferred cross-section reduce the supertranslation freedom to the
translation freedom so that a preferred
Poincar{\'e} subgroup can be picked out from the BMS group.
The same holds in the limits
$u\rightarrow \pm \infty$, in which the requirement
of a finite radiative energy loss implies
$\partial_u \sigma \rightarrow 0$. Although the electric part of
the strain can be gauged away at either $u=+\infty$
or $u=-\infty$, the memory effect $\Delta \sigma$
is gauge invariant and (\ref{eq:super}) determines a 
supertranslation shift
\begin{equation}
    q^A q^B \eth_A \eth_B \,\alpha(x^C) =\Delta \sigma(x^C)
\end{equation}
between the preferred Poincar{\'e} groups
at $u=\pm \infty$. 
The rotation subgroups picked out by the initial and final
preferred Poincar{\'e} groups differ by this supertranslation. 
As a result, the corresponding components of angular momentum
intrinsic to the initial and final states differ by supermomenta.

Only the electric part of the strain is affected by supertranslations
because $\alpha$ is real and $\sigma$ is intrinsically complex.
The decomposition of the strain into electric and magnetic parts
is analogous to the E-mode/B-mode decomposition of electromagnetic waves. 
The magnetic part of the null memory effect
must vanish except for matter fields whose stress-energy
tensor satisfies properties which are not expected for astrophysical systems~\cite{sky}. The exceptional
cases are matter fields whose magnetic part of their stress
is anisotropic and has nonvanishing
time derivative in the limit of infinite future or past
retarded time in the neighbourhood of
${\mathcal I}^+$. Recently, consistent with general physical
principles, a shell of matter with anisoptropic magnetic stress which
expands to timelike infinity in a radial flow has
been constructed which produces ordinary memory
of the magnetic type~\cite{waldmag}. 

The supertranslation shift between the initial and final
preferred  Poincar{\'e} groups complicates the interpretation
of angular momentum flux conservation laws.
This could lead to a distinctly general relativistic mechanism
for angular momentum loss. Although the intermediate radiative 
epoch must be treated by numerical methods, the
Kerr-Schild model developed here provides a
framework for such investigations.


\ack
We are grateful to the AEI in Golm  for hospitality during this project.
We thank I. R{\' acz} for comments on the manuscript
and R. M. Wald for informing us of recent results.
TM appreciates support from the members of the
N{\'u}cleo de Astronom{\'ia} and  Faculty of Engineering of
University Diego Portales, Santiago.  
JW was supported by NSF grants PHY-1505965
and PHY-1806514 to the University of Pittsburgh. 

\end{document}